\documentclass[reprint,showpacs,aps,nofootinbib,superscriptaddress]{revtex4-1}
\usepackage{amssymb}
\usepackage{amsmath,amssymb,graphicx}
\usepackage{graphicx}
\usepackage{dcolumn}
\usepackage{bm}
\usepackage{float}
\usepackage{multirow}

\def\beq{\begin{equation}}
\def\eeq{\end{equation}}
\def\bea{\begin{eqnarray}}
\def\eea{\end{eqnarray}}

\newcommand{\overbar}[1]{\mkern 1.5mu\overline{\mkern-1.5mu#1\mkern-1.5mu}\mkern 1.5mu}

\usepackage{pstricks}
\usepackage{amsbsy}
\usepackage{bm}
\usepackage{color}
\usepackage{fancyhdr}
\usepackage[pdftex]{epsfig}
\usepackage[colorlinks=true,linkcolor=blue,citecolor=blue,urlcolor=magenta,filecolor=blue]{hyperref}
\usepackage{slashed}

\begin{document}

\bigskip

\vspace{2cm}

\title{Search for lepton-number-violating signals in the charm sector}
\vskip 6ex

\author{Diego Milan\'{e}s}
\email{diego.milanes@cern.ch}
\affiliation{Departamento de F\'{i}sica, Universidad Nacional de Colombia, C\'{o}digo Postal 11001, Bogot\'{a}, Colombia}
\author{N\'{e}stor Quintero}
\email{nestor.quintero01@usc.edu.co}
\affiliation{Facultad de Ciencias B\'{a}sicas, Universidad Santiago de Cali, Campus Pampalinda, Calle
5 No. 62-00, C\'{o}digo Postal 76001, Santiago de Cali, Colombia}

\bigskip
\begin{abstract}
We explore signals of lepton-number-violation in the charm physics sector. We study the four-body $|\Delta L|= 2$ decays of the $D^0$ meson, $D^0 \to P^- \pi^-\mu^+ \mu^+$ ($P = \pi, K$) as an alternative evidence of the Majorana nature of neutrinos. We carry out an exploratory study on the potential sensitivity that LHCb experiment could achieve for these $|\Delta L|= 2$ processes. We show that for a long term expected integrated luminosity of 300 fb$^{-1}$, a signal significance of branching ratios of the order $\mathcal{O}(10^{-9})$ might be accessible, allowing to improve the experimental bounds obtained by the E791 experiment. Limits on the parameter space of a heavy sterile neutrino that could be obtained from their experimental search are discussed as well.
\end{abstract}

\maketitle
\bigskip

\section{Introduction}

Looking for lepton-number-violating (LNV) signals in neutrinoless double-$\beta$ ($0\nu\beta\beta$) nuclear decay is considered as the most attractive and sensitive way to prove that neutrinos are their own antiparticles (or not), i.e. elucidate if neutrinos are Majorana particles (or Dirac ones)~\cite{deGouvea:2013,Rodejohann:2011,Vissani:2015,Gomez-Cadenas}. Nowadays, the experiments Majorana, GERDA, CUORE, EXO-200 and KamLAND-Zen~\cite{Aalseth:2017btx,Agostini:2018tnm,Alduino:2017ehq,Albert:2017owj,KamLAND-Zen} have reported the best lower limits on the half-lives of different isotopes ($^{76}{\rm Ge}, ^{136}{\rm Xe},  ^{130}{\rm Te}$) that typically leads to $T_{1/2}^{0\nu} \gtrsim 10^{25}$ yr. Despite all these experimental effort, the lack of evidence of this $|\Delta L| =2$ process opens the possibility of pursuing different low-energy search pathways as alternative evidence to test the Majorana nature of neutrinos.
This complementarity is also reinforced by the fact that a positive observation of $0\nu\beta\beta$ decay can only probe the first fermion family (LNV $ee$ sector), while alternative LNV searches are accessible to different leptonic sectors~\cite{Rodejohann:2011}.

Since their experimental search is accessible to di\-ffe\-rent flavor facilities, the low-energy studies of 	$|\Delta L|= 2$ decays of pseudoscalar mesons $K, D, D_s, B, B_c,B_s$ (both charged and neutral) and the $\tau$ lepton have attracted a lot of theoretical attention~\cite{Atre:2009,Kovalenko:2000,Ali:2001,Atre:2005,Kovalenko:2005,Helo:2011,Cvetic:2010,
Zhang:2011,Bao:2013,Wang:2014,Quintero:2016,Sinha:2016,Gribanov:2001,Quintero:2011,Quintero:2012b,
Quintero:2013,Dong:2013,Yuan:2013,Quintero:2012a,Dib:2012,Dib:2014,Yuan:2017,
Shuve:2016,Asaka:2016,Cvetic:2016,Cvetic:2017,Zamora-Saa:2016,Yuan:2017uyq,Kim:2017pra,Mejia-Guisao:2017gqp,Cvetic:CP}, where different final-state topologies have been considered. 
An interesting way of realizing these $|\Delta L|= 2$ decays is through the exchange of an intermediate on-shell Majorana neutrino $N$ with a kinematically allowed mass (typically, hundreds of MeV up to few GeV), leading to a considerably enhancement of the decay rates~\cite{Atre:2009,Atre:2005,Kovalenko:2005,Helo:2011,Cvetic:2010,
Zhang:2011,Bao:2013,Wang:2014,Quintero:2016,Sinha:2016,Gribanov:2001,Quintero:2011,Quintero:2012b,
Quintero:2013,Dong:2013,Yuan:2013,Quintero:2012a,Dib:2012,Dib:2014,Yuan:2017,Shuve:2016,Asaka:2016,
Cvetic:2016,Cvetic:2017,Zamora-Saa:2016,Yuan:2017uyq,Kim:2017pra,Mejia-Guisao:2017gqp,Cvetic:CP}.
In Refs.~\cite{Drewes:2016lqo,Asaka:2016zib} have found that the $0\nu\beta\beta$ rate can also be enhanced due to the contribution from heavy neutrino exchange with masses in the GeV scale.
Interestingly enough, new physics scenarios with heavy Majorana neutrinos within this GeV mass range have been investigated as a simultaneous explanation to the neutrino mass  generation and the baryon asymmetry of the Universe (via leptogenesis)~\cite{Shaposhnikov:2005,Shaposhnikov:2013,Drewes:2014,GeV_Leptogenesis,Rasmussen:2016}.

\begin{figure}[!t]
\begin{center}
\includegraphics[angle=0,scale=0.44]{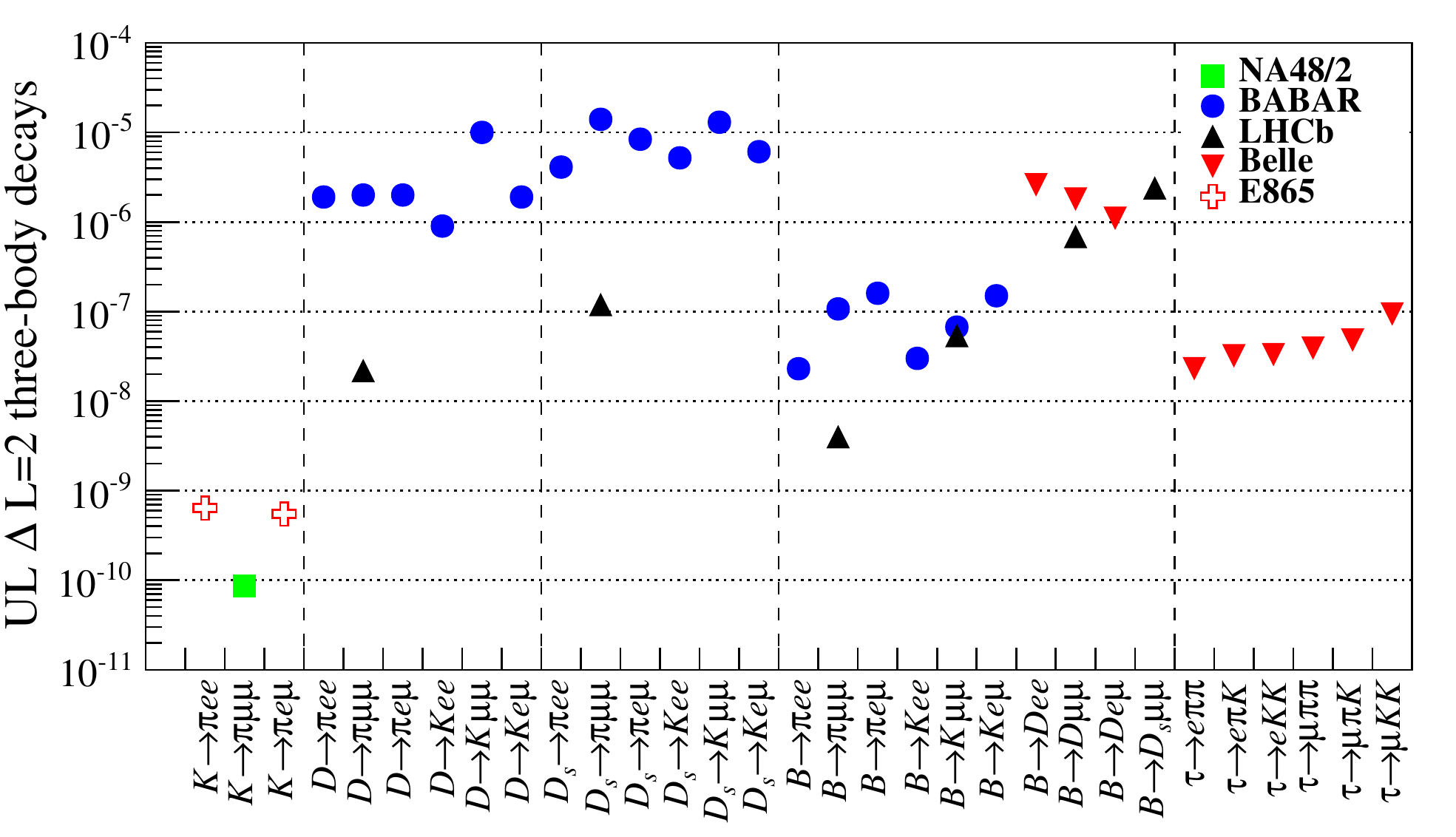}\hfill
\includegraphics[angle=0,scale=0.44]{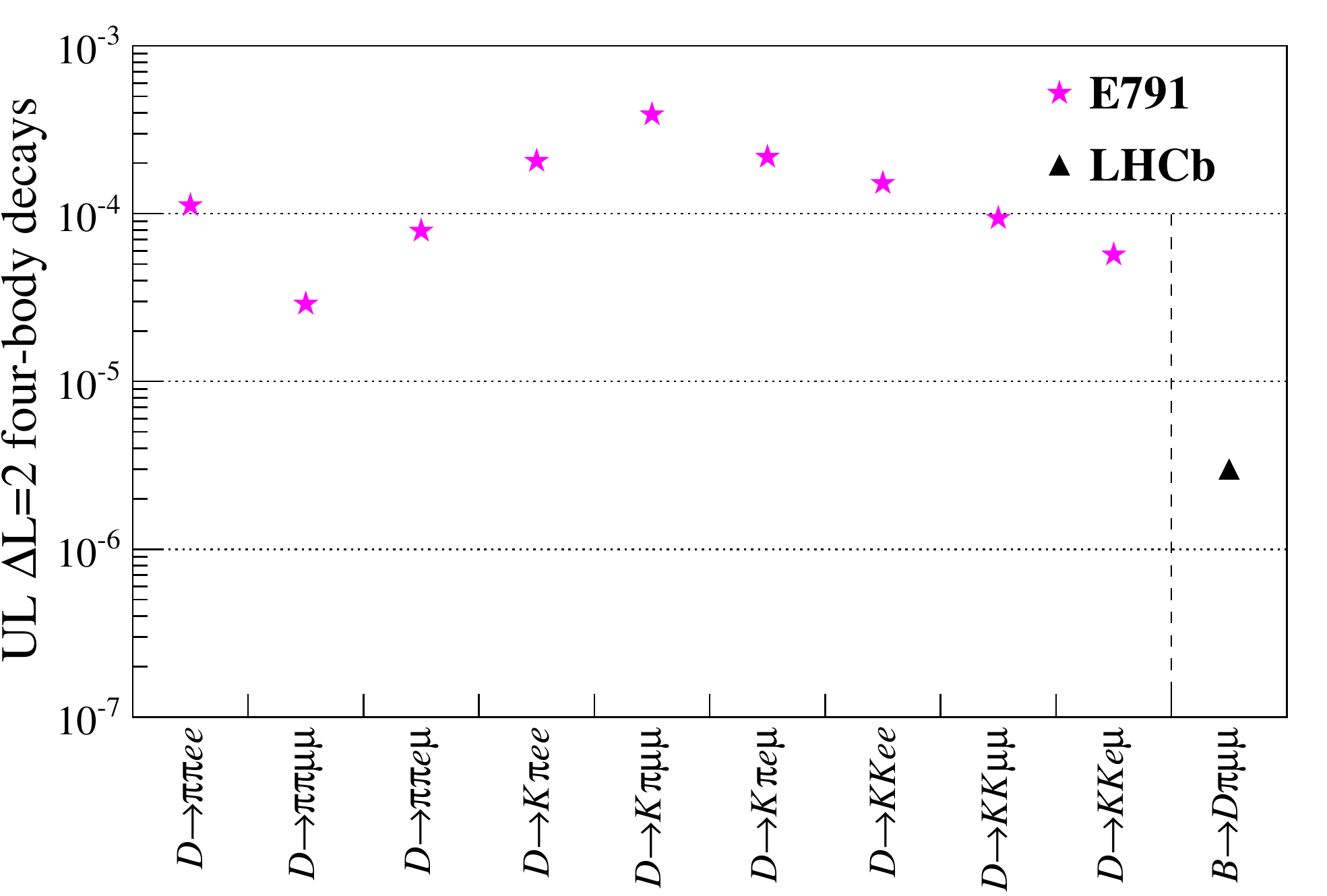}
\caption{\small Current experimental UL on branching ratios of three- (top) and four-body (bottom) $\Delta L=2$ decays. Taken from NA48/2, E865, BABAR, Belle, LHCb, and E791 Collaborations~\cite{CERNNA48/2:2016,Appel:2000tc,BABAR,BABAR:2014,LHCb:2012,LHCb:2013,LHCb:2014,Belle:2011,Belle:2013,
E791}. For simplicity we have not included the UL on $B^- \to V^+ \ell^- \ell^{\prime -}$, with $V=\rho,K^{\ast},D^{\ast}$, which are typically of the order $\lesssim 10^{-7}$~\cite{BABAR:2014}.}
\label{fig:1} 
\end{center}
\end{figure}

Experimentally, upper limits (UL) on the branching ratios of various of these $|\Delta L|= 2$ decays have been obtained by the experiments NA48/2, E865, BABAR, Belle, LHCb, and E791  \cite{CERNNA48/2:2016,Appel:2000tc,BABAR,BABAR:2014,LHCb:2012,LHCb:2013,LHCb:2014,Belle:2011,Belle:2013,E791}. See also the Particle Data Group~\cite{PDG}. In Fig.~\ref{fig:1}, we present a summary of the current UL for different three- and four-body channels. The strongest limits come from kaon sector, particularly from the channel ${\rm BR}(K^- \to \pi^+\mu^-\mu^-) < 8.6 \times10^{-11}$~\cite{CERNNA48/2:2016}; while in the heavy flavor sector, the channel ${\rm BR}(B^- \to \pi^+\mu^-\mu^-) < 4.0 \times10^{-9}$~\cite{LHCb:2014} provides the strongest one. A long term integrated luminosity of 300 fb$^{-1}$ is expected in future LHCb upgrade, and whereas for Belle II, a 50-fold increase in integrated luminosity is expected greater than previous record (Belle and BABAR), allowing to improve the sensitivity on the $|\Delta L|= 2$ channels. Perspectives on the experimental sensitivity of the NA62 experiment to searches of heavy neutrinos has been discussed as well~\cite{Drewes:2018gkc}. Furthermore, recent studies show the sensitivity of the LHCb and CMS experiments to look for LNV signals in $|\Delta L|= 2$ processes of $ \Lambda_b$ baryon and $B_s$ meson~\cite{Mejia-Guisao:2017,Mejia-Guisao:2017gqp}.

Regarding LNV searches in the charm physics, the LHCb collaboration found 90\% confidence level limits on the branching ratios of the three-body decays ${\rm BR}(D^+ \to \pi^-\mu^+\mu^+) < 2.2 \times 10^{-8}$ and ${\rm BR}(D_s^+ \to \pi^-\mu^+\mu^+) < 1.2 \times 10^{-7}$~\cite{LHCb:2013}, that improve by several orders of magnitude the previous limits obtained by BABAR~\cite{BABAR}. While, the four-body channels $D^0 \to h^-h'^-\ell^+\ell^{\prime +}$, where $h,h'=\pi,K$ and $\ell,\ell^{\prime}=e, \mu$, were studied by the E791 collaboration~\cite{E791} more than a decade ago and 90\% confidence level  upper limits of order $\mathcal{O}(10^{-5}-10^{-4})$ for their branching ratios were obtained. Currently, the LHCb experiment has collected the largest sample of charmed mesons and supplementary searches of $|\Delta L|= 2$ channels could be performed, thus complementing previous LHCb analysis~\cite{LHCb:2012,LHCb:2013,LHCb:2014}. 
 
Our aim in this work is to explore a charmed search to track the possible LNV signals at the LHCb experiment by studiying the four-body $|\Delta L|= 2$ decays of the $D^0$ meson, $D^0 \to P^- \pi^-\mu^+ \mu^+$ ($P = \pi, K$). Their search will provide an alternative evidence of the Majorana  nature of neutrinos, allowing to prove the LNV $\mu\mu$ sector. Without referring to any new physics scenario, we will consider a simplified approach in which one heavy Majorana neutrino $N$ couples to a charged lepton ($\ell=\mu$) whose its strength is characterized by the quantity $V_{\ell N}$~\cite{Atre:2009}. We will treat the mass $m_N$ and mixing $V_{\mu N}$ of this heavy sterile neutrino as unknown phenomenological parameters that can be constrained (set) from the experimental non-observation (observation) of $|\Delta L| =2$ processes~\cite{Atre:2009,Helo:2011,Quintero:2016}.  	
We carry out an exploratory study on the potential sensitivity that LHCb experiment could achieve for these $|\Delta L|= 2$ processes (same-sign $\mu^+ \mu^+$), by taking into account its corresponding signal significance. We will show that branching fractions sensitivity at the LHCb experiment will be able to improve by several orders of magnitude the experimental limits obtained by E791~\cite{Quintero:2013,E791}.

The paper is structured as follows. In Sec.~\ref{LNV} we study the four-body LNV decays of the $D^0$ meson. The Sec.~\ref{sensitivity} contains the experimental sensitivity at the LHCb. We present in Sec.~\ref{constraints} the exclusion regions on the parameter space  $(m_N,|V_{\mu N}|^2)$ that can be achieved from their experimental searches at the LHCb experiment. We close with concluding remarks in Sec.~\ref{Conclusion}.

\section{Four-body $|\Delta L| = 2$ decays of $D^0$ meson}  \label{LNV}

We conduct a study of the four-body $|\Delta L|= 2$ decays of the $D^0$ meson, $D^0 \to P^- \pi^-\mu^+ \mu^+$, with $P = \pi, K$ denoting a final-state light meson. Under the simplified assumption that one Majorana neutrino $N$, with a kinematically allowed mass in the range $m_N \in [0.25,1.62]$ GeV such that it can be produced on-shell in these processes, dominates the decay amplitude. These four-body $|\Delta L|= 2$ channels have been previously studied in Refs.~\cite{Quintero:2013,Dong:2013,Yuan:2013} using different approaches for the evaluation of the hadronic transition $D \to P$. Taking the UL reported by E791 experiment~\cite{E791}, in Ref.~\cite{Quintero:2013} obtained bounds on the parameters space of a heavy neutrino ($m_N,|V_{\mu N}|^2$) that turned out to be very mild. Moreover, the authors of Ref.~\cite{Dong:2013} estimated from 2.9 fb$^{-1}$ Monte Carlo sample that BESIII experiment could get an UL on the channel $D^0 \to K^- \pi^- e^+ e^+$ of the order $1 \times 10^{-9}$; nevertheless, such a sensitivity is far below the one obtained from $0\nu\beta\beta$ decay. Here, we present a reanalysis of these $|\Delta L|= 2$ channels by considering the recent lattice QCD calculations of the semileptonic $D \to P$ form factors~\cite{Lubicz:2017syv}. We pay particular attention to the $\mu^+ \mu^+$ channels and their experimental signal at the LHCb experiment (see Sec.~\ref{sensitivity}).

The branching fraction of the four-body $|\Delta L|= 2$ decays $D^0 \to P^- \pi^-\mu^+ \mu^+$ can be written in the factorized form 
\begin{eqnarray}\label{4leptonic}
\mathcal{B}(D^0  \to P^-\pi^-\mu^+\mu^+) &=& \mathcal{B}(D^0 \to P^- \mu^+ N) \nonumber \\
&& \times  \Gamma(N \to \mu^+\pi^-) \tau_N  /\hbar,
\end{eqnarray}  

\noindent  where the on-shell Majorana neutrino is produces through the semileptonic decay $D^0 \to P^- \mu^+N$ and consecutively $N \to \mu^+\pi^-$, with $\tau_N$ as the lifetime of the Majorana neutrino. The decay width of $N \to \mu^+\pi^-$ is given by the expre\-ssi\-on \cite{Atre:2009}
\begin{equation}
\Gamma(N  \to \mu^+\pi^-) = |V_{\mu N}|^2 \ \bar{\Gamma}(N  \to  \mu^-\pi^+),
\end{equation}

\noindent with
\begin{eqnarray}
\bar{\Gamma}(N \to \mu^+\pi^-) &=&  \dfrac{G_F^2}{16 \pi}|V_{ud}^{\text{CKM}}|^2  f_\pi^2 m_N \sqrt{\lambda(m_N^2,m_\mu^2,m_\pi^2)}\nonumber \\
&& \times \bigg[ \bigg(1- \dfrac{m_\mu^2}{m_N^2} \bigg)^2 - \dfrac{m_\pi^2}{m_N^2} \bigg(1+ \dfrac{m_\mu^2}{m_N^2} \bigg) \bigg], \nonumber \\ \label{Ntopimu}
\end{eqnarray}

\noindent where $G_F$ is the Fermi constant, $f_{\pi}$ is the pion decay constant, and $V_{ud}^{\text{CKM}}$ is the Cabbibo-Kobayashi-Maskawa (CKM) matrix element involved. 
	
The branching ratio of $D^0 \to P^-  \mu^+ N$ is given by the expression~\cite{Cvetic:2016} 
\begin{equation}
\mathcal{B}(D^0 \rightarrow P^-  \mu^+ N) =  |U_{\mu N}|^2  \int dt \dfrac{d\overbar{\mathcal{B}}(D^0 \rightarrow P^-  \mu^+ N)}{dt} ,
\end{equation}

\noindent where
\begin{eqnarray} \label{BR_Bs}
&& \dfrac{d\overbar{\mathcal{B}}(D^0 \rightarrow P^-  \mu^+ N)}{dt} \nonumber  \\ 
&& = \dfrac{G_F^2 \tau_{D^0}}{384\pi^3 m_{D}^3 \hbar} |V_{cq}^{\rm CKM}|^2 \ \dfrac{\big[\lambda(m_\mu^2,m_N^2,t)\lambda(m_{D}^2,m_P^2,t) \big]^{1/2}}{t^3} \nonumber \\
&& \times  \Big(\big[F_+^{DP}(t)\big]^2 C_+(t) + \big[F_0^{DP}(t)\big]^2 C_0(t) \Big) ,
\end{eqnarray} 

\noindent is the so-called differential canonical branching ratio~\cite{Cvetic:2016}, where $V_{cq}^{\text{CKM}}$ is the CKM element (with $q = d, s$ for $P=\pi, K,$); and $F_{+}^{DP}(t)$ and $F_{0}^{DP}(t)$ are the vector and scalar form factors for the $D \to P$ transition, respectively, which are eva\-lua\-ted at the square of the transferred momentum $t=(p_{D}-p_P)^2$. The kinematic coefficients $C_+(t)$ and $C_0(t)$ involved in Eq.~\eqref{BR_Bs} are defined as
\begin{eqnarray}
C_+(t) &=& \lambda(m_{D}^2,m_P^2,t) [2t^2 + t(m_\mu^2 + m_N^2) + (m_\mu^2 - m_N^2)^2],\nonumber\\ 
&&  \\
C_0(t) &=& 3(m_{D}^2 - m_P^2) [m_\mu^2(t+2m_N^2 -m_\mu^2) \nonumber\\ 
&&+ m_N^2(t -m_N^2)],
\end{eqnarray}

\noindent respectively, where $\lambda(x,y,z)=x^{2}+y^{2}+z^{2}-2(xy+xz+yz)$ is the usual kinematic K\"{a}llen function. The total branching fraction is then obtained by integrating the differential canonical branching ratio over the full $t$ region $[(m_\mu+m_N)^2$,$(m_{D}-m_P)^2]$.

In later calculations we will use the following inputs~\cite{PDG}: $|V_{ud}^{\text{CKM}}|= 0.97417$, $|V_{cd}^{\text{CKM}}|= 0.218$, $|V_{cs}^{\text{CKM}}|= 0.997$, and $f_{\pi}= 130.2(1.7)$ MeV\footnote{See the review  ``Leptonic decays of charged pseudoscalar mesons" from PDG~\cite{PDG}.}. The masses of particles involved are taken from~\cite{PDG}. For the form factors associated with the $D\to P$ transition, we will use the theoretical predictions provided by the lattice QCD approach~\cite{Lubicz:2017syv}.

\section{Experimental sensitivity at the LHCb}  \label{sensitivity}

The LHCb experiment is a perfect scenario to perform searches for LNV processes from heavy hadron decays, given the excellent detector performance and the large amount of data that has been collected, and that will be collected during future LHC runs ~\cite{Alves:2008zz,Aaij:2014jba,{Bediaga:2012uyd}}. Using an integrated luminosity of 2 fb$^{-1}$ collected from $pp$ collisions at a center-of-mass energy of 8 TeV, the LHCb collaboration observed the decays $D^0\to\pi^+\pi^-\mu^+\mu^-$ and $D^0\to K^+K^-\mu^+\mu^-$~\cite{Aaij:2017iyr}. These decays share the same type of particles in the final state as the LNV mode under study, and therefore we can use information from this analysis to extrapolate sensitivity considerations of $D^0\to$ LNV in the framework of the LHCb experiment for different data sample sizes.

In the LHCb analysis the $D^0$ meson candidates, are extracted from a $D^{*+}\to D^0\pi^+$ sample, produced directly from the $pp$ collision vertex. Given the small phase-space in this decay, there is a clean signature to select the $D^0$ candidates and reduce random background events. If the selected $D^0$ mesons were selected to come directly from the $pp$ collision, the sample of $D^0$ would have been larger, but the background level would have made unfeasible the extraction of the signal.  The measured branching fraction for these decays are $\mathcal{B}_{\pi\pi}\equiv\mathcal{B}(D^0\to\pi^+\pi^-\mu^+\mu^-)=(9.64\pm1.20)\times 10^{-7}$ and $\mathcal{B}_{KK}\equiv\mathcal{B}(D^0\to K^+K^-\mu^+\mu^-)=(1.54\pm0.32)\times 10^{-7}$, where the quoted error contains the statistical and systematic uncertainties. The extracted signal yields, after combining several di-muon regions of study, are $N_{\pi\pi}=561\pm28$ and $N_{KK}=34\pm6$ signal events, as stated in Table 1 of Ref.~\cite{Aaij:2017iyr}, for $D^0\to\pi^+\pi^-\mu^+\mu^-$ and $D^0\to K^+K^-\mu^+\mu^-$ respectively. A conservative approximation is to consider that signal efficiency in the LHCb experiment will remain constant along the years, which is highly unlikely since the detector and algorithms are in constant improvement, thus the number of expected events of a decay with the same particles in the final state, should scale linearly with the luminosity and with the cross-section, which up to a good approximation scales linearly, as well, with the center-of-mass energy. Hence, a good estimation of the number of events of a decay with same final state as the LNV modes under study, for different conditions of luminosity $\mathcal{L}$, $pp$ collisions at a different center-of-mass energy $\sqrt{s}$ and different branching fraction $\mathcal{B}$, is
\begin{equation}
N^{\text{LNV}}_{S,hh}(\mathcal{L},\sqrt{s},\mathcal{B}) = \frac{\mathcal{L}\cdot \sqrt{s}\cdot \mathcal{B}\cdot N_{hh}}{2 \ \text{fb}^{-1}\cdot 8 \ \text{TeV}\cdot \mathcal{B}_{hh}},
\end{equation}
where the subindex $hh$ refers to the different hadronic decays.

\begin{figure}[hbt!]
\begin{center}
\includegraphics[scale=0.67]{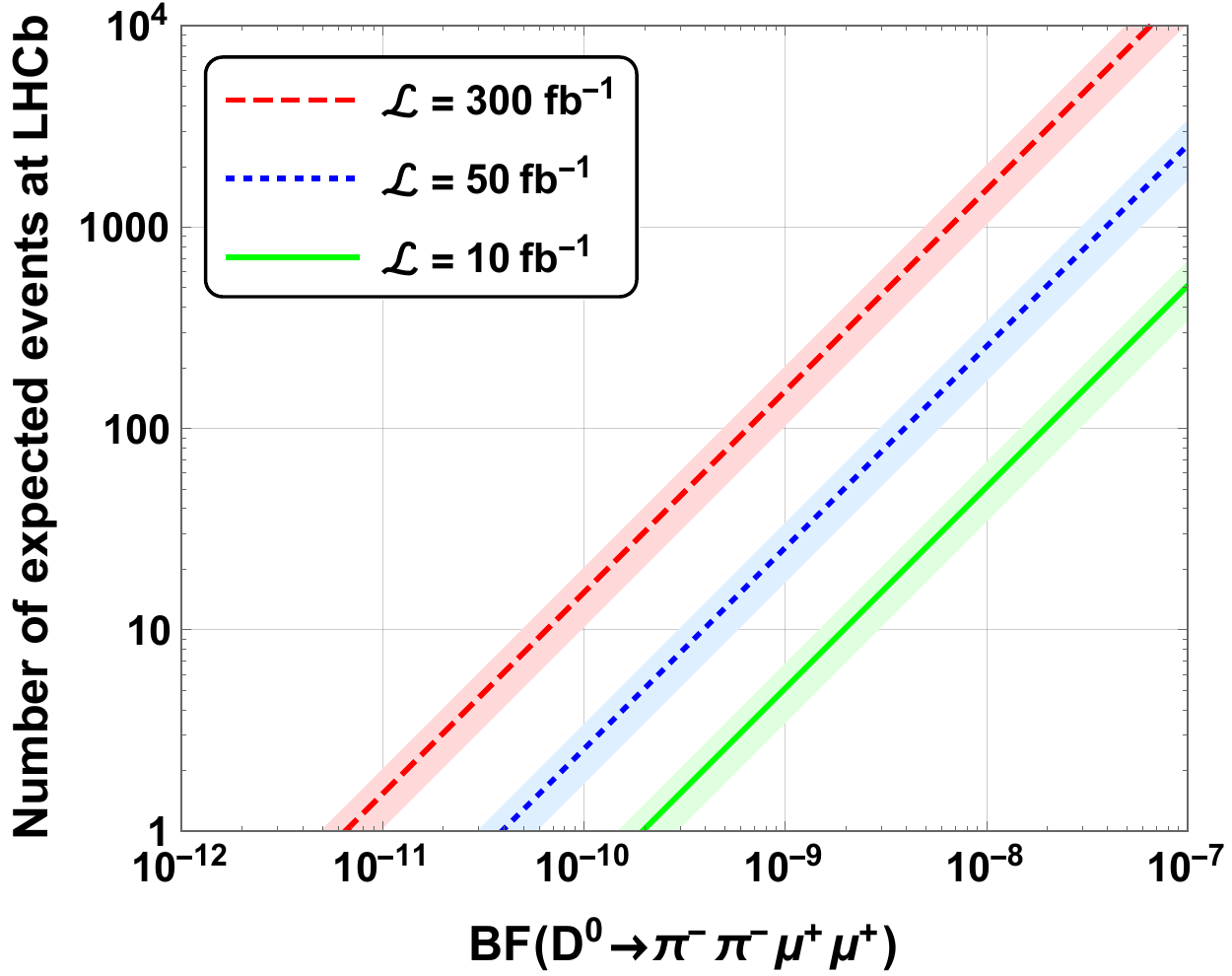}
\includegraphics[scale=0.67]{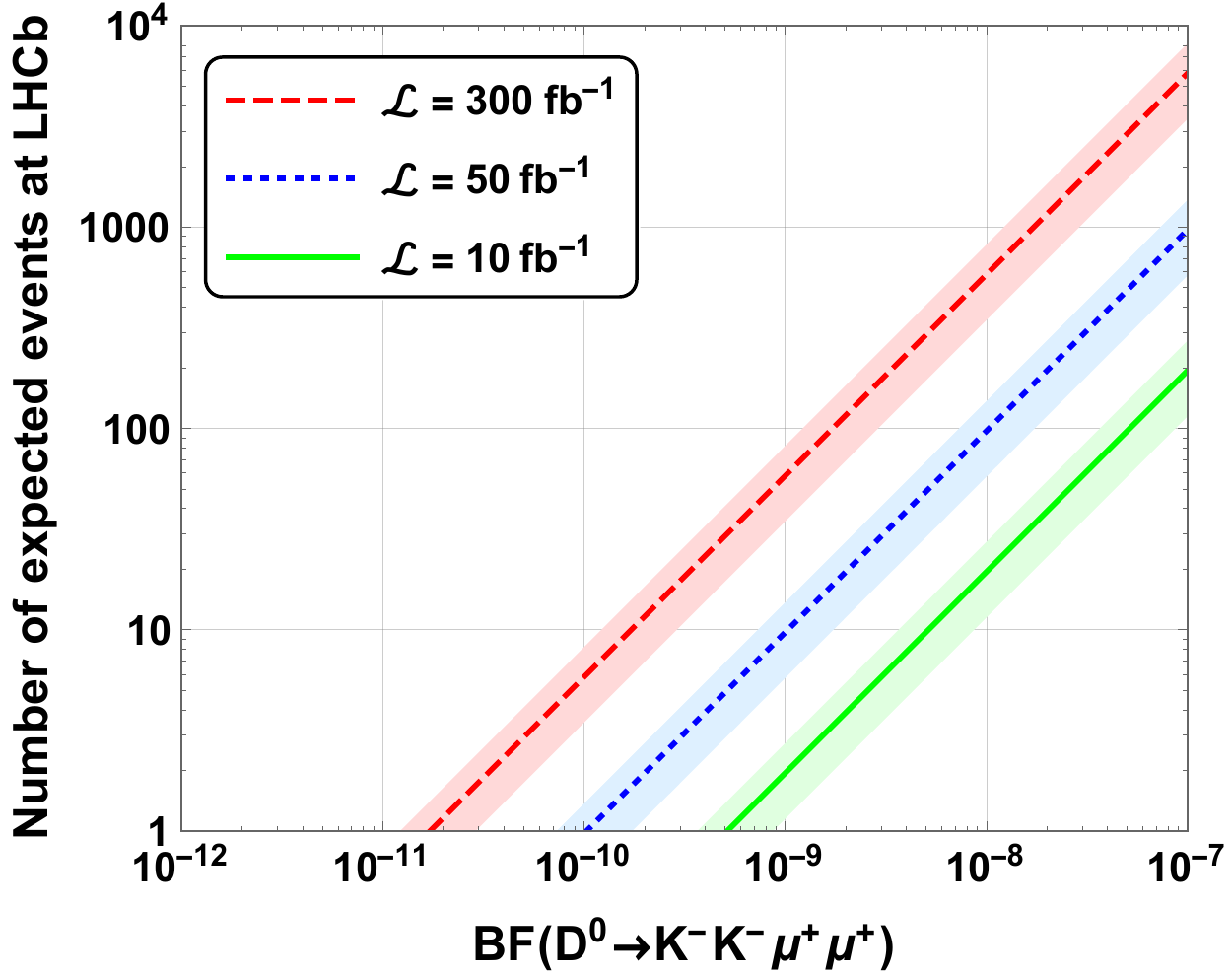}

\caption{Number of expected events at LHCb, for $D^0\to\pi^-\pi^-\mu^+\mu^+$ (top) and $D^0\to K^-K^-\mu^+\mu^+$ (bottom) as a function of the branching fraction, for different luminosity values: $\mathcal{L}=10\ \text{fb}^{-1}$ (solid green), $50 \ \text{fb}^{-1}$ (dotted blue), and $300 \ \text{fb}^{-1}$ (dashed red). The filled region represents the uncertainty in the computation.}
\label{fig:lhcb1}
\end{center}
\end{figure}

We have considered three different LHCb scenarios, $\mathcal{L}$=10, 50, and $300 \ \text{fb}^{-1}$, which correspond to the typical projections of short, middle and long term expected integrated luminosities, respectively. Figure~\ref{fig:lhcb1} shows an estimation of the number of events that can be detected in the LHCb experiment as a function of the branching fraction and integrated luminosity for the two modes. In Ref.~\cite{Aaij:2016xmb} a study of the variation of the reconstruction efficiency, with fully simulated Monte Carlo samples dedicated to the LHCb experiment, is performed for long lived particles with mass within 20 - 80~GeV/$c^2$ and lifetimes in the range of 5 - 100~ps. Hence, the uncertainties shown in Fig.~\ref{fig:lhcb1} correspond to the propagation of the error in the signal yields and in addition a 30\% of uncertainty has been assigned to account for efficiency effects in the reconstruction of massive lived neutrinos. In Table~\ref{tab:yields} the number of expected LNV events in LHCb, for a given value of integrated luminosity and branching fraction is reported, showing that in long term, for values of branching fraction above $10^{-10}$ there will be chances to detect LNV $D^0\to h^-h^-\mu^+\mu^+$ decays.

\begin{table}[htb!]
\renewcommand{\arraystretch}{1.2}
\renewcommand{\arrayrulewidth}{0.8pt}
\begin{center}
\begin{tabular}{cccc}
\hline\hline
 $\mathcal{L}$ (fb$^{-1}$) & $\mathcal{B}$ & $N^{\text{LNV}}_{S,\pi\pi}$ & $N^{\text{LNV}}_{S,KK}$\\
    \hline
    \multirow{3}{*}{10} & $10^{-7}$ & $509\pm167$ & $193\pm78$\\
    					& $10^{-8}$ & $51\pm17$ & $19\pm8$\\
                        & $10^{-9}$ & $5\pm2$ & $2\pm1$\\
    \hline
    \multirow{4}{*}{50} & $10^{-7}$ & $2546\pm836$ & $966\pm392$\\
    					& $10^{-8}$ & $255\pm84$ & $97\pm39$\\
                        & $10^{-9}$ & $26\pm8$ & $10\pm4$\\
                        & $10^{-10}$ & $3\pm1$ & - \\
    \hline
    \multirow{4}{*}{300} & $10^{-7}$ & $15276\pm5016$ & $5795\pm2350$\\
    					 & $10^{-8}$ & $1528\pm502$ & $580\pm235$\\
                         & $10^{-9}$ & $153\pm50$ & $58\pm23$\\
                         & $10^{-10}$ & $15\pm5$ & $6\pm2$\\
\hline\hline
\end{tabular}
\end{center}
\caption{Number of expected LNV events for a given value of integrated luminosity and branching fraction.}
\label{tab:yields}
\end{table}

However, the number of detected events is not always a good indicator of the sensitivity to claim discovery of such type of events. In this case, the signal significance will gave a better estimation of the real chance of observing the LNV $D^0$ decay. In a large-sample limit, the discovery significance $Z$, is given by~\cite{PDG} 
\begin{equation}\label{eq:significance}
Z=\sqrt{2\left((N_S+N_B)\log\frac{N_S+N_B}{N_B} -N_S\right)},
\end{equation}
where $N_S$ and $N_B$ denote the number of signal and background events respectively. In Table 1 of Ref~\cite{Aaij:2017iyr}, not only number of signal events are quoted but also the significance, therefore a background estimation can be done by finding the roots of Eq. \ref{eq:significance}, and after we can extrapolate to our specific energy and luminosity scenario. In the LHCb analysis two main sources of background are treated, random combinatoric and peaking background from misidentified hadrons as muons. Same sources background will be present in the LNV modes and therefore we do not split among those background sources, and instead consider all sources as one. From performing the procedure above mentioned, we found in the  $D^0\to\pi^+\pi^-\mu^+\mu^-$ LHCb sample a about $235\pm15$ background events, and for the  $D^0\to K^+K^-\mu^+\mu^-$
a total of $7\pm3$, Where in both cases the uncertainty is assigned as $\sqrt{N_B}$. Assuming that the background scales with the luminosity and with the center-of-mass energy, just as the signal yield, the extrapolation of background events expected in the LNC decay modes is computed as
\begin{equation}
N^{\text{LNV}}_{B,hh}(\mathcal{L},\sqrt{s}) = \frac{\mathcal{L}\cdot \sqrt{s}\cdot N_{B,hh}}{2 \ \text{fb}^{-1}\cdot 8 \ \text{TeV}}.
\end{equation}

\begin{figure}[hbt!]
\begin{center}
\includegraphics[scale=0.67]{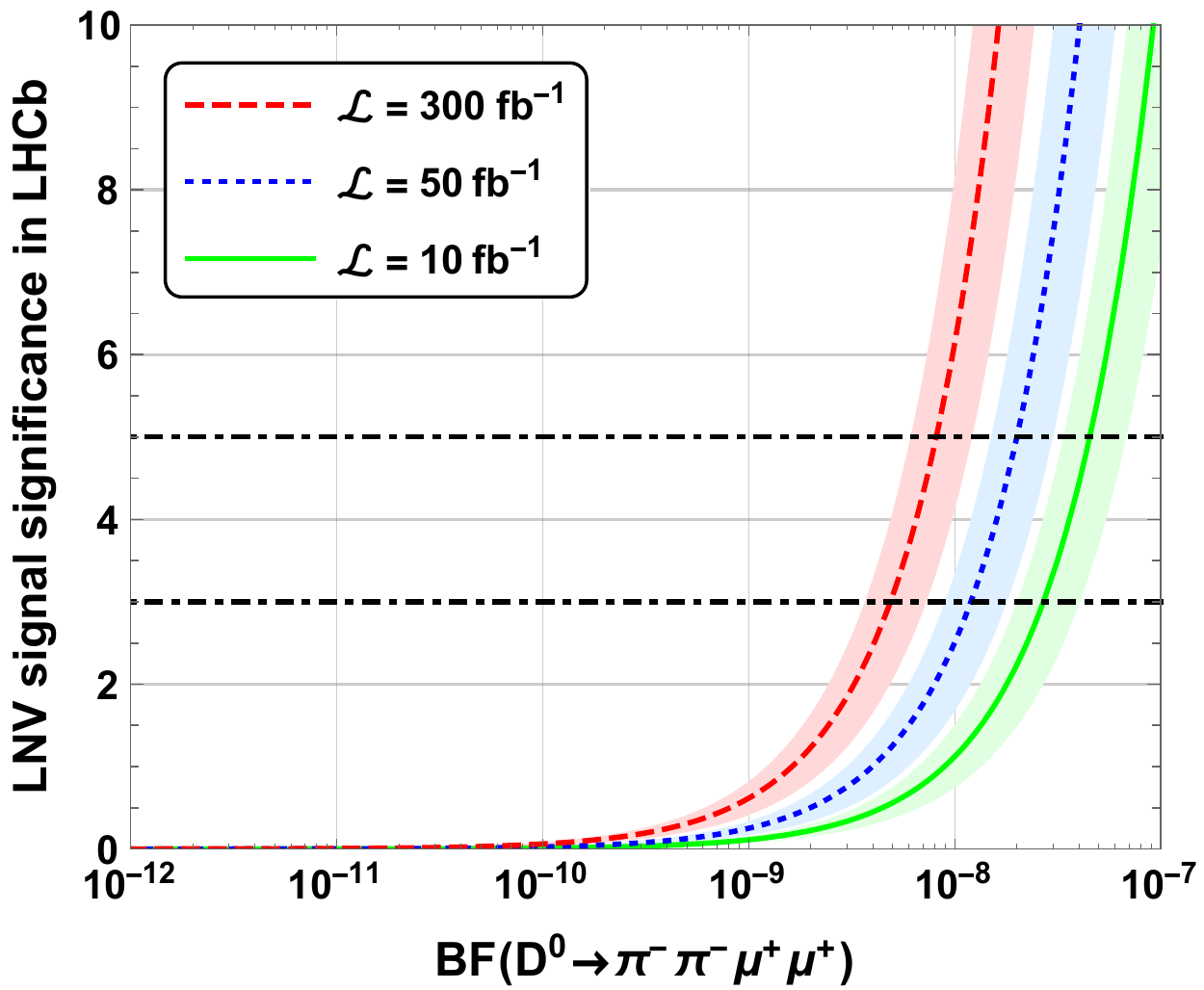}
\includegraphics[scale=0.67]{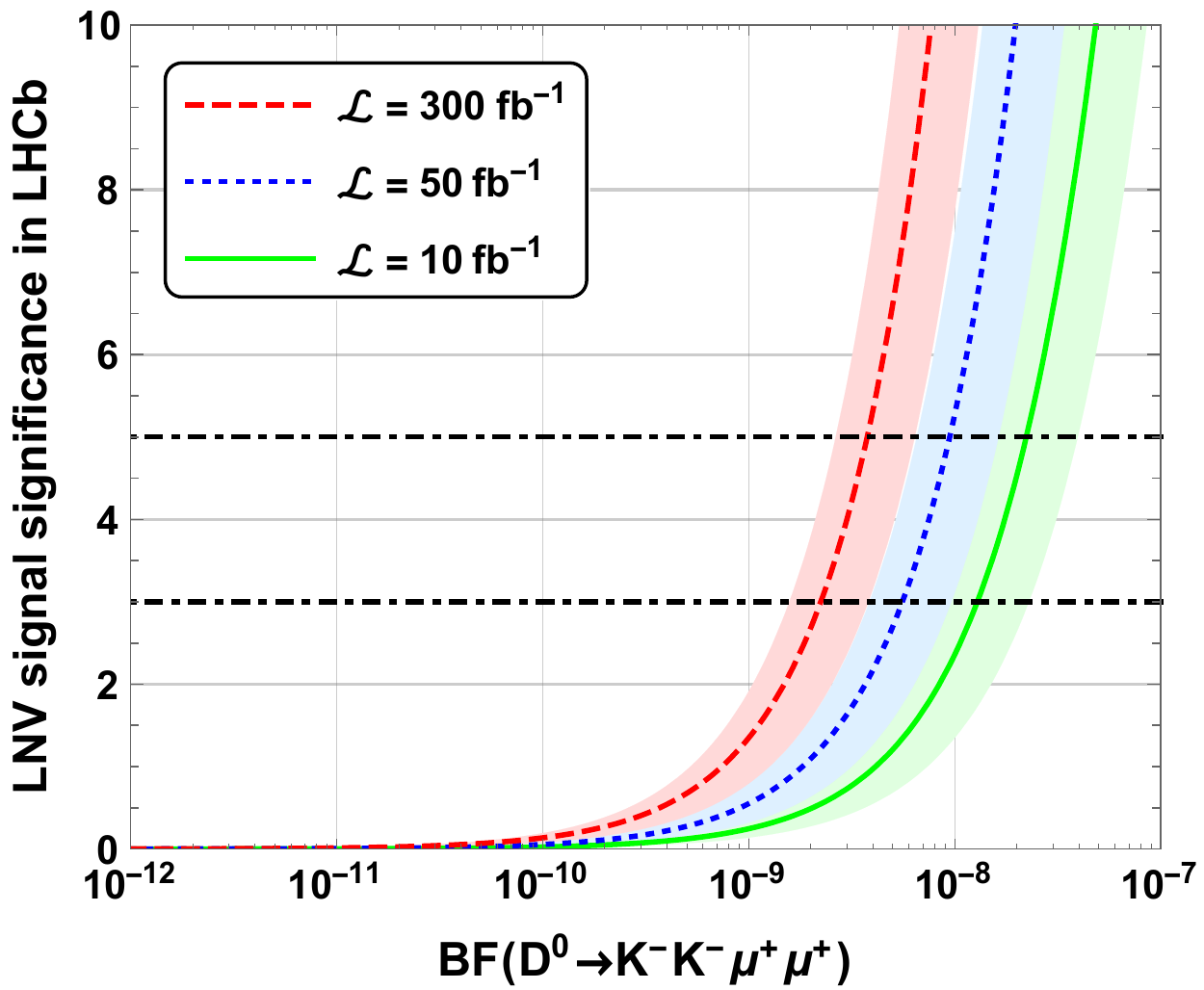}
\caption{Signal significance expected in the LHCb experiment, for $D^0\to\pi^-\pi^-\mu^+\mu^+$ (top) and $D^0\to K^-K^-\mu^+\mu^+$ (bottom) as a function of the branching fraction, for different luminosity values: $\mathcal{L}=10 \ \text{fb}^{-1}$ (solid green), $50 \ \text{fb}^{-1}$ (dotted blue), and $300 \ \text{fb}^{-1}$ (dashed red). The filled region represents the uncertainty in the computation. Horizontal black dot-dashed lines correspond to $3\sigma$ and $5\sigma$ limit.}
\label{fig:lhcb2}
\end{center}
\end{figure}

Figure~\ref{fig:lhcb2} shows how the signal significance changes with the branching fraction of the LNV decays. The extrapolated background level is quoted in Table~\ref{tab:sig}, where it is also quoted the minimum branching fraction of the LNV from which observation can be achieved in the LHCb experiment. This show that for a long term expected integrated luminosity of 300 fb$^{-1}$, branching fractions of the order $\mathcal{O}(10^{-9})$ would be reachable, allowing to improve by several orders of magnitude the experimental limits obtained by E791 experiment.

\begin{table}[htb!]
\renewcommand{\arraystretch}{1.3}
\renewcommand{\arrayrulewidth}{0.8pt}
\begin{center}
\begin{tabular}{ccccc}
\hline\hline
 $\mathcal{L}$ (fb$^{-1}$)  &  $N^{\text{LNV}}_{B,\pi\pi}$ &$\mathcal{B}^{\text{LNV}}_{\pi\pi}$ & $N^{\text{LNV}}_{B,KK}$& $\mathcal{B}^{\text{LNV}}_{KK}$\\
    \hline
10 & 2056 & $>3.4\times10^{-8}$ & 61 & $>1.5\times10^{-8}$\\
50 & 10281& $>1.5\times10^{-8}$& 306 &$>6.6\times10^{-9}$ \\
300 & 61687& $>6.1\times10^{-9}$ & 1837 &$>2.7\times10^{-9}$ \\
\hline\hline
\end{tabular}
\end{center}
\caption{Extrapolated background level, and branching fraction from which there can be observation of LNV modes at LHCb}
\label{tab:sig}
\end{table}


\section{Exclusion regions on the parameter space $(m_N,|V_{\mu N}|^2)$}  \label{constraints}

\begin{figure*}[!t]
\centering
\includegraphics[angle=0,scale=0.44]{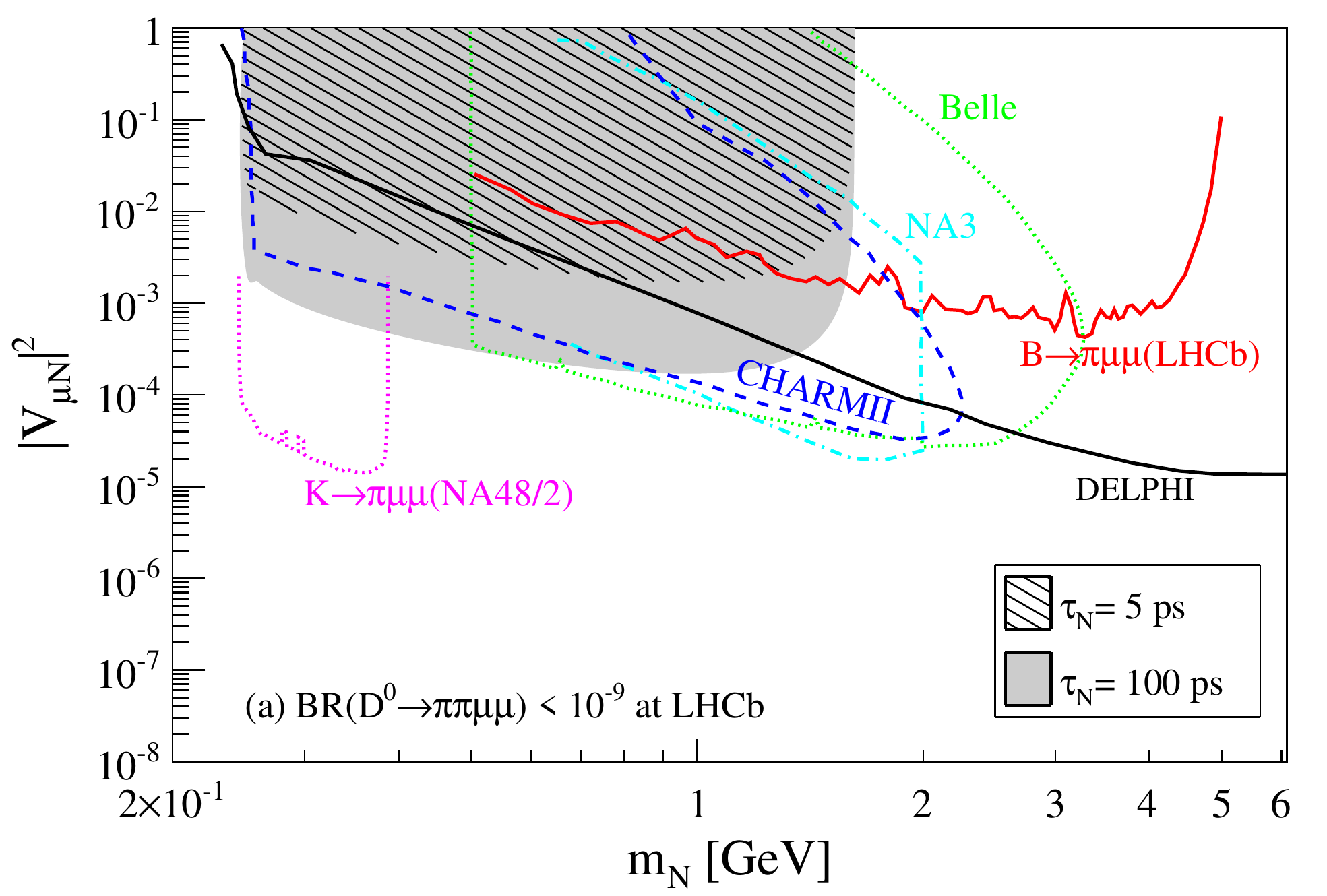}
\includegraphics[angle=0,scale=0.44]{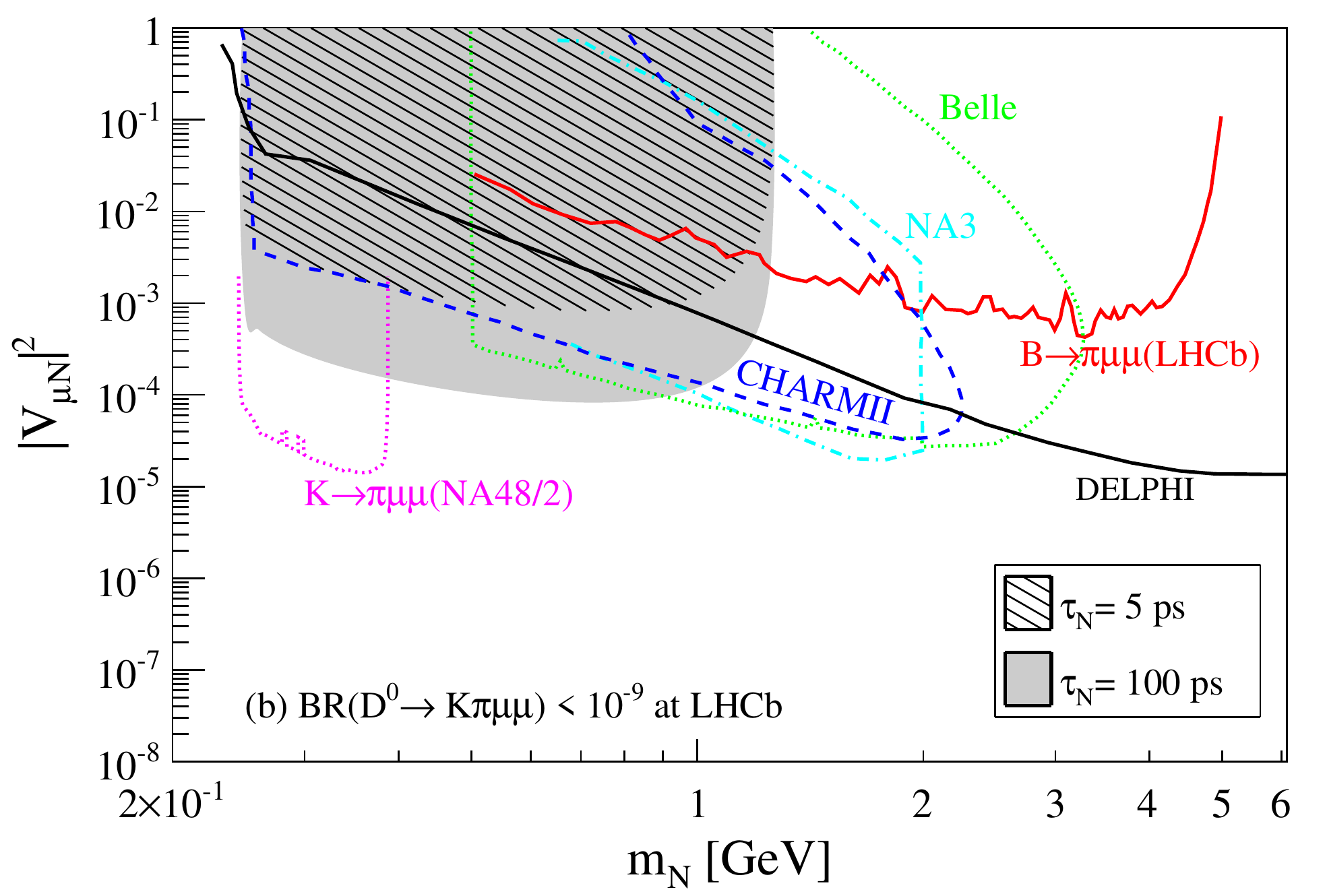}
\caption{\small Exclusion regions on the $(m_N, |V_{\mu N}|^{2})$ plane for (a) ${\rm BR}(D^0 \to  \pi^- \pi^-\mu^+ \mu^+) < 10^{-9}$ and  (b) ${\rm BR}(D^0 \to  K^- \pi^-\mu^+ \mu^+) < 10^{-9}$ at LHCb. The black, blue, and gray regions represent the bounds obtained for heavy neutrino lifetimes of $\tau_N$ = 5 and 100 ps, respectively. Limits provided by $K^- \rightarrow \pi^{+}\mu^{-}\mu^{-}$~\cite{CERNNA48/2:2016}, $B^{-} \to \pi^{+}\mu^{-}\mu^{-}$~\cite{Shuve:2016}, Belle~\cite{Belle:N}, DELPHI~\cite{LEP}, NA3~\cite{NA3}, and CHARMII~\cite{CHARMII}, are also included for comparison.}
\label{Fig4.1}
\end{figure*}

Based on the LHCb sensitivity analysis presented in the previous Sec. \ref{sensitivity}, in the following, we examine the bounds on the parameter space of a heavy sterile neutrino $(m_N,|V_{\mu N}|^2)$ that can be achieved from the experimental searches on $D^0 \to  (\pi^-\pi^-,K^-\pi^-)\mu^+ \mu^+$ at the LHCb experiment. 

To constraint the squared magnitude $|V_{\mu N}|^2$ as a function of the mass  $m_N$, the following relation obtained from Eq.~\eqref{4leptonic} can be used for that purpose
\begin{equation}
|V_{\mu N}|^2 = \Bigg[  \frac{\mathcal{B}(D^0 \to  P^- \pi^-\mu^+ \mu^+) \ \hbar}{ \overbar{\mathcal{B}}(D^0 \to  P^- \mu^+ N) \times \overbar{\Gamma}(N \to \mu^+\pi^-) \tau_N}\Bigg]^{1/2} ,
\end{equation}

\noindent where $\overbar{\mathcal{B}}(D^0 \to  P^-\mu^+N)$ and $\overbar{\Gamma}(N \to \mu^+\pi^-)$ are given by Eqs.~\eqref{BR_Bs} and~\eqref{Ntopimu}, respectively. We will consider heavy neutrino lifetimes of $\tau_N = [5, 100]$ ps as benchmark points in our analysis. This will allow us to extract limits on $|V_{\mu N}|^2$ without any additional assumption on the relative size of the mixing matrix elements.

Considering an expected LHCb sensitivity on the branching fractions of the order $\mathcal{B}(D^0 \to  P^- \pi^-\mu^+ \mu^+) < 10^{-9}$ for 300 fb$^{-1}$, in Figs.~\ref{Fig4.1}(a) and~\ref{Fig4.1}(b) we show the exclusions regions on the $(m_N,|V_{\mu N}|^2)$ plane obtained from future searches on $D^0 \to  \pi^- \pi^-\mu^+ \mu^+$ and $D^0 \to  K^- \pi^-\mu^+ \mu^+$, respectively.  In both cases, the black, blue, and gray regions represent the bounds obtained for heavy neutrino lifetimes of $\tau_N$ = 5 and 100 ps, respectively. We also plot the exclusion limits obtained from searches on $|\Delta L|=2$ channels, $K^- \rightarrow \pi^{+}\mu^{-}\mu^{-}$ from NA48/2 (taken for $\tau_N =$ 1000 ps)~\cite{CERNNA48/2:2016} and $B^{-} \to \pi^{+}\mu^{-}\mu^{-}$ from LHCb~\cite{LHCb:2014}, for comparison. For the $B^{-} \to \pi^{+}\mu^{-}\mu^{-}$ channel, we compare with the revised limit \cite{Shuve:2016} from the LHCb analysis~\cite{LHCb:2014}. In the range of $m_N$ relevant for $D^0 \to  (\pi^- \pi^-, K^- \pi^-) \mu^+ \mu^+$ channels, we can observe that the most stringent bound is given by $K^- \to \pi^+ \mu^-\mu^-$, which can reach $|V_{\mu N}|^2\sim \mathcal{O}(10^{-5})$, but only for a very narrow mass window of [0.25, 0.38] GeV. For $m_N > 0.38$ GeV, the four-body cha\-nnels under study would improve the region of $|V_{\mu N}|^{2}$ covered by the channel $B^- \to \pi^+\mu^-\mu^-$.

For further comparison, in Figs.~\ref{Fig4.1}(a) and~\ref{Fig4.1}(b) we also show the limits coming from different search strategies such as Belle~\cite{Belle:N}, DELPHI~\cite{LEP}, NA3~\cite{NA3}, and CHARMII~\cite{CHARMII} experiments\footnote{For recent reviews on the theoretical and experimental status of different GeV-scale heavy neutrino search strategies see for instance Refs.~\cite{Bondarenko:2018ptm,Drewes:2015,Deppisch:2015,deGouvea:2015,Fernandez-Martinez:2016} and references therein.}. As can be seen, our $|\Delta L|=2$ channels proposals have comparable sensitivity that different search strategies in the region mass where these overlap. In particular, searches on $D^0 \to  K^- \pi^- \mu^+ \mu^+$ could slightly improve those limits in the mass window of $\sim$ [0.38, 1] GeV.

We encourage the experimental colleagues from the LHCb experiment to look for heavy Majorana neutrinos through the search on four-body $|\Delta L|= 2$ decays of the $D^0$ meson, thus complementing previous LHCb analysis~\cite{LHCb:2012,LHCb:2013,LHCb:2014}. Additionally, these searches can be also performed at the Belle II and BESIII experiments.

\section{Conclusions} \label{Conclusion}

We have explored a charmed search to track the possible signals of lepton-number-violation at the LHCb experiment, due to the copious charm production. We studied the four-body $|\Delta L|= 2$ decays of the $D^0$ meson, $D^0 \to P^- \pi^-\mu^+ \mu^+$ ($P = \pi, K$), induced by an on-shell Majorana neutrino $N$ with a mass of few GeV. We performed an exploratory study on the potential sensitivity (signal significance) that LHCb experiment could achieve for these $|\Delta L|= 2$ processes. For a long term expected integrated luminosity of 300 fb$^{-1}$, we found that branching fractions of the order $\mathcal{O}(10^{-9})$ might be feasible. Such a sensitivity will improve by several orders of magnitude the experimental limits obtained by E791. It is also found that for a neutrino mass window of $0.25 \leq m_N \leq 1.62$ GeV, exclusion regions on the parameter space $(m_N,|V_{\mu N}|^2)$ that could be obtained from their experimental search will have comparable sensitivity that previous bounds. Particularly, searches on $D^0 \to  K^- \pi^- \mu^+ \mu^+$ could slightly improve bounds in the range of $\sim$ [0.38, 1] GeV.


\acknowledgments

We are grateful with Jhovanny Mej\'{i}a-Guisao and Jos\'{e} D. Ruiz-\'{A}lvarez for their collaboration at the early stage of this work. N. Quintero acknowledges support from Direcci\'{o}n General de Investigaciones - Universidad Santiago de Cali and the hospitality from Instituto de F\'{i}sica - Universidad de Antioquia, where the completion of this work was done.




\begin{thebibliography}{99}

	\bibitem{deGouvea:2013}
A. de Gouv\^{e}a and P. Vogel, Lepton flavor and number conservation, and physics beyond the standard model, Prog. Part. Nucl. Phys. \textbf{71}, 75 (2013) \href{http://arxiv.org/abs/1303.4097}{\texttt{[arXiv:1303.4097 [hep-ph]]}}.

\bibitem{Rodejohann:2011} 
H. P\"{a}s and W. Rodejohann, Neutrinoless Double Beta Decay, New J. Phys. \textbf{17}, 115010 (2015), \href{http://arxiv.org/abs/1507.00170}{\texttt{[arXiv:1507.00170 [hep-ph]]}};
W.~Rodejohann, Neutrinoless Double Beta Decay and Particle Physics,
Int. J. Mod. Phys. E \textbf{20}, 1833 (2011)
\href{http://arxiv.org/abs/1106.1334}{\texttt{[arXiv:1106.1334]}}.

\bibitem{Vissani:2015}
S. Dell'Oro, S. Marcocci, M. Viel, and F. Vissani, Neutrinoless double beta decay: 2015 review, Adv. High Energy Phys. \textbf{2016}, 2162659 (2016) \href{http://arxiv.org/abs/1601.07512}{\texttt{[arXiv:1601.07512 [hep-ph]]}}.     
    
\bibitem{Gomez-Cadenas}
J. J. G\'{o}mez-Cadenas \textit{et al}, The search for neutrinoless double beta decay, Riv. Nuovo Cim. \textbf{35}, 29 (2012) \href{http://arxiv.org/abs/1109.5515}{\texttt{[arXiv:1109.5515 [hep-ph]]}}.

  

\bibitem{Aalseth:2017btx} 
C.~E.~Aalseth {\it et al.} (Majorana Collaboration), Search for Neutrinoless Double-$\beta$ Decay in $^{76}$Ge with the Majorana Demonstrator, Phys.\ Rev.\ Lett.\  {\bf 120}, 132502 (2018) \href{http://arxiv.org/abs/1710.11608}{\texttt{[arXiv:1710.11608 [nucl-ex]]}}. 
  
\bibitem{Agostini:2018tnm} 
M.~Agostini {\it et al.} (GERDA Collaboration), Improved Limit on Neutrinoless Double-$\beta$ Decay of $^{76}$Ge from GERDA Phase II, Phys.\ Rev.\ Lett.\  {\bf 120}, 132503 (2018) \href{http://arxiv.org/abs/1803.11100}{\texttt{[arXiv:1803.11100 [nucl-ex]]}}.

\bibitem{Alduino:2017ehq} 
C.~Alduino {\it et al.} (CUORE Collaboration), First Results from CUORE: A Search for Lepton Number Violation via $0\nu\beta\beta$ Decay of $^{130}$Te, Phys.\ Rev.\ Lett.\  {\bf 120}, 132501 (2018) \href{http://arxiv.org/abs/1710.07988}{\texttt{[arXiv:1710.07988 [nucl-ex]]}}.

\bibitem{Albert:2017owj} 
 J.~B.~Albert {\it et al.} (EXO Collaboration), Search for Neutrinoless Double-Beta Decay with the Upgraded EXO-200 Detector, Phys.\ Rev.\ Lett.\  {\bf 120}, 072701 (2018) \href{http://arxiv.org/abs/1707.08707}{\texttt{[arXiv:1707.08707 [nucl-ex]]}}.

\bibitem{KamLAND-Zen}
A. Gando \textit{et al.} (KamLAND-Zen Collaboration), Search for Majorana Neutrinos near the Inverted Mass Hierarchy Region with KamLAND-Zen, Phys. Rev. Lett. \textbf{117}, 082503 (2016)
\href{http://arxiv.org/abs/1605.02889}{\texttt{[arXiv:1605.02889 [hep-ex]]}}.

 

\bibitem{Atre:2009}
A. Atre, T. Han, S. Pascoli, and B. Zhang, The search for heavy Majorana neutrinos, 
J. High Energy Phys. \textbf{05}, 030 (2009)
\href{http://arxiv.org/abs/0901.3589}{\texttt{[arXiv:0901.3589 [hep-ph]]}}.

\bibitem{Kovalenko:2000}
L. S. Littenberg and R. Shrock, Upper Bounds on Lepton Number Violating Meson Decays,
Phys. Rev. Lett. \textbf{68}, 443 (1992);
C. Dib, V. Gribanov, S. Kovalenko, and I. Schmidt, K meson neutrinoless double muon
decay as a probe of neutrino masses and mixings, Phys. Lett. B \textbf{493}, 82 (2000) \href{http://arxiv.org/abs/hep-ph/0006277}{\texttt{[hep-ph/0006277]}}; L. S. Littenberg and R. Shrock, Implications of improved upper bounds on $|\Delta L| = 2$ processes, Phys. Lett. B \textbf{491}, 285 (2000) \href{http://arxiv.org/abs/hep-ph/0005285}{\texttt{[hep-ph/0005285]}};
K. Zuber, New limits on effective Majorana neutrino masses from rare kaon decays, Phys. Lett. B \textbf{479}, 33 (2000) \href{http://arxiv.org/abs/hep-ph/0003160}{\texttt{[hep-ph/0003160]}}.

\bibitem{Ali:2001}
A. Ali, A. V. Borisov, and N. B. Zamorin, Majorana neutrinos and same-sign dilepton
production at LHC and in rare meson decays, Eur. Phys. J. C \textbf{21}, 123 (2001)
\href{http://arxiv.org/abs/hep-ph/0104123}{\texttt{[hep-ph/0104123]}}.

\bibitem{Atre:2005}
A. Atre, V. Barger, and T. Han, Upper bounds on lepton-number violating processes, Phys. Rev. D \textbf{71}, 113014 (2005)  \href{http://arxiv.org/abs/hep-ph/0502163}{\texttt{[hep-ph/0502163]}}.

\bibitem{Kovalenko:2005}
M. A. Ivanov, S. G. Kovalenko, Hadronic structure aspects of $K^+ \to \pi^- + l^+_1 + l^+_2$ decays.
Phys. Rev. D \textbf{71}, 053004 (2005) \href{http://arxiv.org/abs/hep-ph/0412198}{\texttt{[hep-ph/0412198]}}.

\bibitem{Helo:2011}
J. C. Helo, S. Kovalenko, and I. Schmidt, Sterile neutrinos in lepton number and lepton flavor violating decays, Nucl. Phys, \textbf{B853}, 80 (2011) \href{http://arxiv.org/abs/1005.1607}{\texttt{[arXiv:1005.1607 [hep-ph]]}}.

\bibitem{Cvetic:2010}
G. Cvetic, C. Dib, S. K. Kang, and C. S. Kim, Probing Majorana neutrinos in rare $K$ and $D$, $D_s$, $B$, $B_c$ meson decays, Phys. Rev. D \textbf{82}, 053010 (2010) \href{http://inspirehep.net/search?p=find+eprint+1005.4282}{\texttt{[arXiv:1005.4282  [hep-ph]]}}.

\bibitem{Zhang:2011}
J. M. Zhang and G. L. Wang, Lepton-number violating decays of heavy mesons, Eur. Phys. J. C \textbf{71}, 1715 (2011)
\href{http://arxiv.org/abs/1003.5570}{\texttt{[arXiv:1003.5570 [hep-ph]]}}.
 
\bibitem{Bao:2013}
S.-S. Bao, H.-L. Li, Z.-G. Si, and Y.-B. Yang, Search for Majorana Neutrino Signal in $B_c$ Meson Rare Decay, Commun. Theor. Phys. \textbf{59}, 472 (2013) \href{http://arxiv.org/abs/1208.5136}{\texttt{[arXiv:1208.5136 [hep-ph]]}}.

\bibitem{Wang:2014}
Y. Wang, S.-S. Bao, Z.-H. Li, N. Zhu, and Z.-G. Si, Study Majorana Neutrino Contribution to $B$-meson Semi-leptonic Rare Decays, Phys. Lett. B \textbf{736}, 428 (2014) \href{http://arxiv.org/abs/1407.2468}{\texttt{[arXiv:1407.2468  [hep-ph]]}}.

\bibitem{Quintero:2016} 
D. Milan\'{e}s, N. Quintero, and C. E. Vera, Sensitivity to Majorana neutrinos in $\Delta L=2$ decays of $B_c$ meson at LHCb, Phys. Rev. D \textbf{93}, 094026 (2016) \href{http://arxiv.org/abs/1604.03177}{\texttt{[arXiv:1604.03177 [hep-ph]]}}.

\bibitem{Sinha:2016} 	
S. Mandal and N. Sinha, Favoured $B_c$ decay modes to search for a Majorana neutrino, Phys. Rev. D \textbf{94}, 033001 (2016) \href{http://arxiv.org/abs/1602.09112}{\texttt{[arXiv:1602.09112 [hep-ph]]}}.

\bibitem{Gribanov:2001}
V. Gribanov, S. Kovalenko and I. Schmidt, Sterile neutrinos in $\tau$ lepton decays,
Nucl. Phys. \textbf{B607}, 355 (2001) 
\href{http://arxiv.org/abs/hep-ph/0102155}{\texttt{[hep-ph/0102155]}}.


\bibitem{Quintero:2011}
D. Delepine, G. L\'opez Castro, and N. Quintero,  Lepton number violation in top quark and neutral $B$ meson decays, Phys. Rev. D \textbf{84}, 096011 (2011) [\textit{ibid} D \textbf{86}, 079905(E) (2012)] 
\href{http://arxiv.org/abs/1108.6009}{\texttt{[arXiv:1108.6009 [hep-ph]]}}.

\bibitem{Quintero:2012a}
G. L\'opez Castro and N. Quintero, Lepton number violation in tau lepton decays,
Nucl. Phys. B Proc. Suppl. \textbf{253-255}, 12 (2014) \href{http://arxiv.org/abs/1212.0037}{\texttt{[arXiv:1212.0037 [hep-ph]]}}.

\bibitem{Quintero:2013}
G. L\'opez Castro and N. Quintero, Bounding resonant Majorana neutrinos from four-body $B$ and $D$ decays, Phys. Rev. D \textbf{87}, 077901 (2013) 
\href{http://arxiv.org/abs/1302.1504}{\texttt{[arXiv:1302.1504 [hep-ph]]}}.

\bibitem{Dong:2013} 	
H.-R. Dong, F. Feng, and H.-B. Li, Lepton number violation in $D$ meson decay, Chin. Phys. C \textbf{39}, 013101 (2015) \href{http://arxiv.org/abs/1305.3820}{[\texttt{arXiv:1305.3820 [hep-ph]]}}.

\bibitem{Yuan:2013} 		
H. Yuan, T. Wang, G.-L. Wang, W.-L. Ju, and J.-M. Zhang, Lepton-number violating four-body decays of heavy mesons, J. High Energy Phys. \textbf{08}, 066 (2013) 
\href{http://arxiv.org/abs/1304.3810}{\texttt{[arXiv:1304.3810 [hep-ph]]}}.

\bibitem{Quintero:2012b}
G. L\'opez Castro and N. Quintero,  Lepton-number-violating four-body tau lepton decays, Phys. Rev. D \textbf{85}, 076006 (2012) [\textit{ibid} D \textbf{86}, 079904(E) (2012)] 
\href{http://arxiv.org/abs/1203.0537}{\texttt{[arXiv:1203.0537 [hep-ph]]}}.

\bibitem{Dib:2012}
C. Dib, J. C. Helo, M. Hirsch, S. Kovalenko, and I. Schmidt, Heavy sterile neutrinos in tau decays and the MiniBooNE anomaly, Phys. Rev. D \textbf{85}, 011301(R) (2012)
\href{http://arxiv.org/abs/1110.5400}{\texttt{[arXiv:1110.5400 [hep-ph]]}}.

\bibitem{Dib:2014} 
C. Dib and C. S. Kim, Remarks on the lifetime of sterile neutrinos and the effect on detection
of rare meson decays $M^{+} \to M^{\prime -} \ell^+\ell^+$,  Phys. Rev. D \textbf{89}, 077301 (2014)
\href{http://arxiv.org/abs/1403.1985}{\texttt{[arXiv:1403.1985 [hep-ph]]}}.

\bibitem{Yuan:2017} 
 H.~Yuan, Y.~Jiang, T.~Wang, Q.~Li and G.-L.~Wang, Lepton Number Violating Four-body Tau Decay, \href{http://arxiv.org/abs/1702.04555}{\texttt{[arXiv:1702.04555 [hep-ph]]}}.
  
\bibitem{Shuve:2016} 
B.~Shuve and M.~E.~Peskin, Revision of the LHCb Limit on Majorana Neutrinos,   Phys.\ Rev.\ D {\bf 94}, 113007 (2016)   \href{http://arxiv.org/abs/1607.04258}{\texttt{[arXiv:1607.04258 [hep-ph]]}}.

\bibitem{Asaka:2016} 
T.~Asaka and H.~Ishida, Lepton number violation by heavy Majorana neutrino in $B$ decays, Phys.\ Lett.\ B {\bf 763}, 393 (2016) \href{http://arxiv.org/abs/1609.06113}{\texttt{[arXiv:1609.06113  [hep-ph]]}}.

\bibitem{Cvetic:2016} 
G.~Cvetic and C.~S.~Kim, Rare decays of $B$ mesons via on-shell sterile neutrinos, Phys.\ Rev.\ D {\bf 94}, 053001 (2016)  [\textit{ibid} D {\bf 95}, 039901(E) (2017)]
\href{http://arxiv.org/abs/1606.04140}{\texttt{[arXiv:1606.04140 [hep-ph]]}}.

\bibitem{Cvetic:2017} 
G.~Cvetic and C.~S.~Kim, Sensitivity limits on heavy-light mixing $|U_{\mu N}|^2$ from lepton number violating $B$ meson decays,   Phys.\ Rev.\ D {\bf 96}, 035025 (2017) \href{http://arxiv.org/abs/1705.09403}{\texttt{[arXiv:1705.09403 [hep-ph]]}}.
  
\bibitem{Zamora-Saa:2016} 
G.~Moreno and J.~Zamora-Saa, Rare meson decays with three pairs of quasi-degenerate heavy neutrinos, Phys.\ Rev.\ D {\bf 94}, 093005 (2016) \href{http://arxiv.org/abs/1606.08820}{\texttt{[arXiv:1606.08820 [hep-ph]]}};  J.~Zamora-Saa, Resonant CP violation in rare tau decay,  J. High Energ. Phys. \textbf{05}, 110 (2017) \href{http://arxiv.org/abs/1612.07656}{\texttt{[arXiv:1612.07656 [hep-ph]]}}.

\bibitem{Yuan:2017uyq}
H.~Yuan, T.~Wang, Y.~Jiang, Q.~Li and G.~L.~Wang, Four-body decays of $B$ meson with lepton number violation,  J.\ Phys.\ G {\bf 45}, 065002 (2018) \href{http://arxiv.org/abs/1710.03886}{\texttt{[arXiv:1710.03886 [hep-ph]]}}.

\bibitem{Kim:2017pra} 
C.~S.~Kim, G.~L\'{o}pez Castro and D.~Sahoo, Discovering intermediate mass sterile neutrinos through $\tau^- \to \pi^- \mu^- e^+ \nu$ (or $\bar{\nu}$) decay, Phys.\ Rev.\ D {\bf 96}, 075016 (2017) \href{http://arxiv.org/abs/1708.00802}{\texttt{[arXiv:1708.00802 [hep-ph]]}}.

\bibitem{Mejia-Guisao:2017gqp} 
 J.~Mej\'{i}a-Guisao, D.~Milan\'{e}s, N.~Quintero and J.~D.~Ruiz-\'{A}lvarez, Lepton number violation in $B_s$ meson decays induced by an on-shell Majorana neutrino, Phys.\ Rev.\ D {\bf 97}, 075018 (2018) \href{http://arxiv.org/abs/1708.01516}{\texttt{[arXiv:1708.01516 [hep-ph]]}}.
  
\bibitem{Cvetic:CP}
G. Cvetic, C. Dib, C. S. Kim, and J. Zamora-Sa\'{a}, Probing the Majorana neutrinos and their CP violation in decays of charged scalar mesons $\pi, K, D, D_s, B, B_c$, Symmetry \textbf{7}, 726 (2015) \href{http://arxiv.org/abs/1503.01358}{\texttt{[arXiv:1503.01358 [hep-ph]]}}; G. Cvetic, C. S. Kim, and J. Zamora-Sa\'{a}, CP violation in lepton number violating semihadronic decays of $K, D, D_s, B, B_c$, Phys. Rev. D \textbf{89}, 093012 (2014) \href{http://arxiv.org/abs/1403.2555}{\texttt{[arXiv:1403.2555 [hep-ph]]}}; CP violation in $\pi^{\pm}$ meson decay, J. Phys. G \textbf{41}, 075004 (2014) \href{http://arxiv.org/abs/1311.7554}{\texttt{[arXiv:1311.7554 [hep-ph]]}}; C. Dib, M. Campos, and C. S. Kim, CP Violation with Majorana neutrinos in $K$ Meson Decays, J. High Energy Phys. \textbf{02}, 108 (2015) \href{http://arxiv.org/abs/1403.8009}{\texttt{[arXiv:1403.8009 [hep-ph]]}}.
  
\bibitem{Drewes:2016lqo}
M.~Drewes and S.~Eijima, Neutrinoless double $\beta$ decay and low scale leptogenesis, Phys.\ Lett.\ B {\bf 763}, 72 (2016) \href{http://arxiv.org/abs/1606.06221}{\texttt{[arXiv:1606.06221 [hep-ph]]}}.

\bibitem{Asaka:2016zib} 
 T.~Asaka, S.~Eijima and H.~Ishida, On neutrinoless double beta decay in the $\nu$MSM, Phys.\ Lett.\ B {\bf 762}, 371 (2016)   \href{http://arxiv.org/abs/1606.06686}{\texttt{[arXiv:1606.06686 [hep-ph]]}}.  
  
\bibitem{Shaposhnikov:2005}
T. Asaka, S. Blanchet, and M. Shaposhnikov, The $\nu$MSM, dark matter and neutrino masses, Phys. Lett. B \textbf{631}, 151 (2005) \href{http://arxiv.org/abs/hep-ph/0503065}{\texttt{[hep-ph/0503065]}};
T. Asaka and M. Shaposhnikov, The $\nu$MSM, dark matter and baryon asymmetry of the universe, Phys. Lett. B \textbf{620}, 17 (2005) \href{http://arxiv.org/abs/hep-ph/0505013}{\texttt{[hep-ph/0505013]}}.

\bibitem{Shaposhnikov:2013}
L. Canetti, M. Drewes, and M. Shaposhnikov, Sterile Neutrinos as the Origin of Dark and Baryonic Matter, Phys. Rev. Lett. \textbf{110}, 061801 (2013)
\href{http://arxiv.org/abs/1204.3902}{\texttt{[arXiv:1204.3902 [hep-ph]]}};
L. Canetti, M. Drewes, T. Frossard, and M. Shaposhnikov, Dark Matter, Baryogenesis and Neutrino Oscillations from Right Handed Neutrinos, Phys. Rev. D \textbf{87}, 093006 (2013)
\href{http://arxiv.org/abs/1208.4607}{\texttt{[arXiv:1208.4607 [hep-ph]]}}. 

\bibitem{Drewes:2014}
L. Canetti, M. Drewes, and B. Garbrecht, Probing leptogenesis with GeV-scale sterile neutrinos at LHCb and Belle II, Phys. Rev. D \textbf{90}, 125005 (2014) 
\href{http://arxiv.org/abs/1404.7114}{\texttt{[arXiv:1404.7114 [hep-ph]]}}.

\bibitem{GeV_Leptogenesis}
M.~Drewes, B.~Garbrecht, D.~Gueter and J.~Klaric, Testing the low scale seesaw and leptogenesis, J. High Energy Phys. {\bf 08}, 018 (2017) \href{http://arxiv.org/abs/1609.09069}{\texttt{[arXiv:1609.09069 [hep-ph]]}};  P.~Hern\'{a}ndez, M.~Kekic, J.~L\'{o}pez-Pav\'{o}n, J.~Racker and J.~Salvado, Testable Baryogenesis in Seesaw Models, J. High Energy Phys. {\bf 08}, 157 (2016)  \href{http://arxiv.org/abs/1606.06719}{\texttt{[arXiv:1606.06719 [hep-ph]]}}; P.~Hern\'{a}ndez, M.~Kekic, J.~L\'{o}pez-Pav\'{o}n, J.~Racker and N.~Rius, Leptogenesis in GeV scale seesaw models, J. High Energy Phys. {\bf 10}, 067 (2015)  \href{http://arxiv.org/abs/1508.03676}{\texttt{[arXiv:1508.03676  [hep-ph]]}}; B.~Shuve and I.~Yavin, Baryogenesis through Neutrino Oscillations: A Unified Perspective,  Phys.\ Rev.\ D {\bf 89}, no. 7, 075014 (2014) \href{http://arxiv.org/abs/1401.2459}{\texttt{[arXiv:1401.2459 [hep-ph]]}}.

\bibitem{Rasmussen:2016} 
 R.~W.~Rasmussen and W.~Winter, Perspectives for tests of neutrino mass generation at the GeV scale: Experimental reach versus theoretical predictions, Phys.\ Rev.\ D {\bf 94}, 073004 (2016) \href{http://arxiv.org/abs/1607.07880}{\texttt{[arXiv:1607.07880 [hep-ph]]}}.
 

\bibitem{CERNNA48/2:2016}
J.~R.~Batley {\it et al.} (NA48/2 Collaboration), Searches for lepton number violation and resonances in $K^{\pm}\to\pi\mu\mu$ decays,  Phys. Lett. B \textbf{769}, 67 (2017) \href{http://arxiv.org/abs/1612.04723}{\texttt{[arXiv:1612.04723 [hep-ex]]}}.

\bibitem{Appel:2000tc} 
R.~Appel {\it et al.} (E865 Collaboration), Search for lepton flavor violation in $K^+$ decays, Phys.\ Rev.\ Lett.\  {\bf 85}, 2877 (2000) \href{http://arxiv.org/abs/hep-ex/0006003}{\texttt{[hep-ex/0006003]}}.
  
\bibitem{BABAR}
J. P. Lees {\it et al.} (BABAR Collaboration), Searches for Rare or Forbidden Semileptonic Charm Decays, Phys. Rev. D \textbf{84}, 072006 (2011) 
\href{http://arxiv.org/abs/1107.4465}{\texttt{[arXiv:1107.4465 [hep-ex]]}}; 
Search for lepton-number violating processes in $B^+ \to h^- l^+ l^+$ decays, Phys. Rev. D \textbf{85}, 071103(R) (2012) 
\href{http://arxiv.org/abs/1202.3650}{\texttt{[arXiv:1202.3650 [hep-ex]]}}.

\bibitem{BABAR:2014}
J. P. Lees {\it et al.} (BABAR Collaboration), Search for lepton-number violating $B^+ \to X^- \ell^+ \ell^{\prime +}$ decays, Phys. Rev. D \textbf{89}, 011102(R) (2014) 
\href{http://arxiv.org/abs/1310.8238}{\texttt{[arXiv:1310.8238 [hep-ex]]}}.

\bibitem{LHCb:2012} 
R. Aaij \textit{et al.}, (LHCb Collaboration), Search for the lepton number violating decays $B^+ \to \pi^- \mu^+\mu^+$ and $B^+ \to K^-\mu^+\mu^+$, Phys. Rev. Lett. \textbf{108}, 101601 (2012) \href{http://arxiv.org/abs/1110.0730 }{\texttt{[arXiv:1110.0730 [hep-ex]]}};
Searches for Majorana neutrinos in $B^-$ decays, Phys. Rev. D \textbf{85}, 112004 (2012) \href{http://arxiv.org/abs/1201.5600}{\texttt{[arXiv:1201.5600 [hep-ex]]}}.

\bibitem{LHCb:2013} 
R. Aaij \textit{et al.}, (LHCb Collaboration), Search for $D^+_{(s)} \to \pi^+ \mu^+ \mu^-$ and $D^+_{(s)} \to \pi^- \mu^+ \mu^+$ decays, Phys. Lett. B \textbf{724}, 203 (2013) 
\href{http://arxiv.org/abs/1304.6365}{\texttt{[arXiv:1304.6365 [hep-ex]]}}. 

\bibitem{LHCb:2014} 
R. Aaij \textit{et al.}, (LHCb Collaboration),  Search for Majorana neutrinos in $B^- \to \pi^+ \mu^- \mu^-$ decays, Phys. Rev. Lett. \textbf{112}, 131802 (2014) 
\href{http://arxiv.org/abs/1401.5361}{\texttt{[arXiv:1401.5361]}}. 

\bibitem{Belle:2011}
O. Seon {\it et al.} (Belle Collaboration), Search for lepton-number-violating $B^+ \to D^- \ell^+ \ell^{\prime +}$ decays, Phys. Rev. D {\bf 84}, 071106(R) (2011) \href{http://arxiv.org/abs/1107.0642}{\texttt{[arXiv:1107.0642]}}.

\bibitem{Belle:2013}
Y. Miyazaki {\it et al.} (Belle Collaboration), Search for lepton-flavor and lepton-number-violating $\tau \to \ell h h^\prime$ decay modes, Phys. Lett. \textbf{B} 719, 346 (2013) \href{http://arxiv.org/abs/1206.5595}{\texttt{[arXiv:1206.5595]}}.

\bibitem{E791} 
E. Aitala {\it et al.} (E791 Collaboration), Search for rare and forbidden charm meson decays $D^0 \to V \ell^+\ell^-$ and $hh\ell\ell$, Phys. Rev. Lett. \textbf{86}, 3696 (2001)
\href{http://arxiv.org/abs/hep-ex/0011077}{\texttt{[hep-ex/0011077]}}.

\bibitem{PDG}
M. Tanabashi \textit{et al.} (Particle Data Group), The Review of Particle Physics, Phys. Rev. D \textbf{98}, 030001 (2018) \href{http://pdg.lbl.gov}{\texttt{[http://pdg.lbl.gov]}}.

\bibitem{Drewes:2018gkc} 
M.~Drewes, J.~Hajer, J.~Klaric and G.~Lanfranchi, NA62 sensitivity to heavy neutral leptons in the low scale seesaw model,  JHEP {\bf 07}, 105 (2018) \href{http://arxiv.org/abs/1801.04207}{\texttt{[arXiv:1801.04207 [hep-ph]]}}.
  
\bibitem{Mejia-Guisao:2017} 
J.~Mej\'{i}a-Guisao, D.~Milan\'{e}s, N.~Quintero and J.~D.~Ruiz-\'{A}lvarez, Exploring GeV-scale Majorana neutrinos in lepton-number-violating $\Lambda_b^0$ baryon decays, Phys.\ Rev.\ D {\bf 96}, 015039 (2017) \href{http://arxiv.org/abs/1705.10606}{\texttt{[arXiv:1705.10606 [hep-ph]]}}.

\bibitem{Lubicz:2017syv} 
V.~Lubicz {\it et al.} (ETM Collaboration), Scalar and vector form factors of $D \to \pi(K) \ell \nu$ decays with $N_f=2+1+1$ twisted fermions, Phys.\ Rev.\ D {\bf 96}, 054514 (2017)
\href{http://arxiv.org/abs/1706.03017}{\texttt{[arXiv:1706.03017 [hep-lat]]}}.

\bibitem{Alves:2008zz} 
A.~A.~Alves, Jr. {\it et al.} [LHCb Collaboration],The LHCb Detector at the LHC, JINST {\bf 3}, S08005 (2008).
  
\bibitem{Aaij:2014jba} 
R.~Aaij {\it et al.} [LHCb Collaboration], LHCb Detector Performance, Int.\ J.\ Mod.\ Phys.\ A {\bf 30}, no. 07, 1530022 (2015)
  \href{arXiv:1412.6352}{[arXiv:1412.6352 [hep-ex]]}.
  
\bibitem{Bediaga:2012uyd} 
  I.~Bediaga {\it et al.} [LHCb Collaboration], Framework TDR for the LHCb Upgrade : Technical Design Report, CERN-LHCC-2012-007, LHCb-TDR-12; The LHCb Collaboration [LHCb Collaboration], Letter of Intent for the LHCb Upgrade, CERN-LHCC-2011-001.
  
\bibitem{Aaij:2017iyr} 
  R.~Aaij {\it et al.} [LHCb Collaboration], Observation of $D^0$ meson decays to $\pi^+\pi^-\mu^+\mu^-$ and $K^+K^-\mu^+\mu^-$ final states, Phys.\ Rev.\ Lett.\  {\bf 119}, 181805 (2017)
  \href{arXiv:1707.08377}{[arXiv:1707.08377 [hep-ex]]}.
  
\bibitem{Aaij:2016xmb} 
R.~Aaij {\it et al.} (LHCb Collaboration), Search for massive long-lived particles decaying semileptonically in the LHCb detector, Eur.\ Phys.\ J.\ C {\bf 77}, 224 (2017)
\href{http://arxiv.org/abs/1612.00945}{\texttt{[arXiv:1612.00945 [hep-ex]]}}.
 
 
\bibitem{Bondarenko:2018ptm} 
K.~Bondarenko, A.~Boyarsky, D.~Gorbunov and O.~Ruchayskiy, Phenomenology of GeV-scale Heavy Neutral Leptons, \href{http://arxiv.org/abs/1502.00477}{\texttt{arXiv:1805.08567 [hep-ph]}}.
  
\bibitem{Drewes:2015}
M. Drewes and B. Garbrecht, Combining experimental and cosmological constraints on heavy neutrinos, Nucl. Phys. \textbf{B921}, 250 (2017) \href{http://arxiv.org/abs/1502.00477}{\texttt{[arXiv:1502.00477 [hep-ph]]}}.

\bibitem{Deppisch:2015}
F. F. Deppisch, P. S. Bhupal Dev, and A. Pilaftsis, Neutrinos and Collider Physics, New J. Phys. \textbf{17}, 075019 (2015) \href{http://arxiv.org/abs/1502.06541}{\texttt{[arXiv:1502.06541]}}.

\bibitem{deGouvea:2015}
A. de Gouv\^{e}a, and A. Kobach, Global constraints on a heavy neutrino, Phys. Rev. D \textbf{93}, 033005 (2016) \href{http://arxiv.org/abs/1511.00683}{\texttt{[arXiv:1511.00683]}}.

\bibitem{Fernandez-Martinez:2016} 
  E.~Fernandez-Martinez, J.~Hernandez-Garcia and J.~Lopez-Pavon, Global constraints on heavy neutrino mixing,   J. High Energy Phys. {\bf 08}, 033 (2016)  \href{http://arxiv.org/abs/1605.08774}{\texttt{[arXiv:1605.08774]}}.

\bibitem{Belle:N}
D. Liventsev \textit{et al.} (Belle Collaboration), Search for heavy neutrinos at Belle, Phys. Rev. D \textbf{87}, 071102(R) (2013) Erratum:[Phys. Rev. D {\bf 95}, 099903 (2017)] \href{http://arxiv.org/abs/1301.1105}{\texttt{[arXiv:1301.1105 [hep-ex]]}}.

\bibitem{LEP}
P. Abreu {\it et al}. (DELPHI Collaboration), Search for neutral heavy leptons produced in Z decays, Z. Phys. {\bf C74}, 57 (1997).

\bibitem{NA3}
J. Badier \textit{et al.} (NA3 Collaboration), Mass and lifetime limits on new longlived particles in 300 GeV/c $\pi^-$ interactions, Z. Phys. C \textbf{31}, 21 (1986).

\bibitem{CHARMII}
P. Vilain \textit{et al.} (CHARM II Collaboration), Search for heavy isosinglet neutrinos, Phys. Lett. B \textbf{343}, 453 (1995); Phys. Lett. B \textbf{351}, 387 (1995).

\bibitem{NuTeV}
A. Vaitaitis \textit{et al.} (NuTeV Collaboration), Search for neutral heavy leptons in a high-energy neutrino beam, Phys. Rev. Lett. \textbf{83}, 4943 (1999) 
\href{http://arxiv.org/abs/hep-ex/9908011}{\texttt{[hep-ex/9908011]}}.

\end{thebibliography}
\end{document}